\def   \ni {\noindent}

\def   \ssk {\vskip  5truept}

\def   \bsk {\vskip 15truept}

\def   \newline {\hfil\break}

\documentstyle[epsfig]{article}
\begin{document}

%
\def\la{\mathrel{\mathchoice {\vcenter{\offinterlineskip\halign{\hfil
$\displaystyle##$\hfil\cr<\cr\sim\cr}}}
{\vcenter{\offinterlineskip\halign{\hfil$\textstyle##$\hfil\cr
<\cr\sim\cr}}}
{\vcenter{\offinterlineskip\halign{\hfil$\scriptstyle##$\hfil\cr
<\cr\sim\cr}}}
{\vcenter{\offinterlineskip\halign{\hfil$\scriptscriptstyle##$\hfil\cr
<\cr\sim\cr}}}}}
\def\ga{\mathrel{\mathchoice {\vcenter{\offinterlineskip\halign{\hfil
$\displaystyle##$\hfil\cr>\cr\sim\cr}}}
{\vcenter{\offinterlineskip\halign{\hfil$\textstyle##$\hfil\cr
>\cr\sim\cr}}}
{\vcenter{\offinterlineskip\halign{\hfil$\scriptstyle##$\hfil\cr
>\cr\sim\cr}}}
{\vcenter{\offinterlineskip\halign{\hfil$\scriptscriptstyle##$\hfil\cr
>\cr\sim\cr}}}}}
\def\degr{\hbox{$^\circ$}}
\def\arcmin{\hbox{$^\prime$}}
\def\arcsec{\hbox{$^{\prime\prime}$}}

\hsize 5truein
\vsize 8truein
\font\abstract=cmr8
\font\keywords=cmr8
\font\caption=cmr8
\font\references=cmr8
\font\text=cmr10
\font\affiliation=cmssi10
\font\author=cmss10
\font\mc=cmss8
\font\title=cmssbx10 scaled\magstep2
\font\alcit=cmti7 scaled\magstephalf
\font\alcin=cmr6 
\font\ita=cmti8
\font\mma=cmr8
\def\ref{\par\noindent\hangindent 15pt}
\null


\title{\ni 
The effect of a spatially varying Galactic spectral index on the maximum
entropy reconstruction of Planck Surveyor satellite data
}                                               

\bsk \bsk
\author{\ni 
A. W. Jones, M.P. Hobson, P. Mukherjee and A.N. Lasenby
}                                                       
\bsk
\affiliation{ 
Cavendish Astrophysics, Madingley Road, Cambridge, CB3 OHE, UK
}                                                
\bsk
\baselineskip = 12pt

\abstract{ABSTRACT \ni
We study the effect of Galactic foregrounds
with spatially varying spectral indices on analysis of simulated data from the
Planck Satellite. We also briefly mention the effect the
extra-galactic point sources have on the data analysis and summarise
the most recent constraints on Galactic emission at GHz frequencies. 
}                                                    
\bsk
\baselineskip = 12pt

\bsk
\baselineskip = 12pt


\text{\ni 1. INTRODUCTION
\ssk
\ni     
The data obtained through observations of the cosmic microwave
background (CMB) in the frequency range of 30 to 900~GHz will be
contaminated with various foregrounds. These foregrounds must be
subtracted before accurate maps of the CMB can be obtained. 

A previous
paper (Hobson, Jones, Lasenby \& Bouchet 1998a, hereafter HJLB98) used a
maximum entropy technique to analyse  
simulated data from the Planck Surveyor satellite that included
six different components of emission in $10\times 10$-degree fields (Bouchet {\sl
et al.} 1997, Gispert \& Bouchet 1997). 
It was found that faithful
reconstructions of five components (CMB, thermal
Sunyaev-Zel'dovich effect and Galactic free-free, synchrotron and
dust emissions) were possible using only frequency spectral
information and faithful reconstructions of all six were possible when
spatial information was included (the above five and the kinetic SZ
which has the same frequency spectrum as the CMB). 
A recent update to that paper (Hobson
{\sl et al.} 1998b) also includes the effects of extra-galactic point
sources in the data (Toffalatti {\sl et al.} 1998) 
and it was found that the reconstructions were not
significantly altered. The analysis technique did not
subtract the effect of the point sources from individual data pixels
but rather treated them as an additional noise term in the covariance
matrix. 

This proceeding updates the Galactic foreground simulations to
relax the assumptions made about the frequency spectra of the Galactic
components and includes spatially varying spectral indices.

}

{\text 
  \ni 2. THE SIMULATIONS AND RECENT UPDATES

HJLB98 presented the simulations used to create the data for the
Planck Satellite observations. Here we use the same input maps so that
easy comparison can be made and we do not include the effects of point
sources as this was found to have little effect on the
reconstructions. Figure 1 shows the six input components at
300~GHz.

\begin{figure}
\centerline{\psfig{file=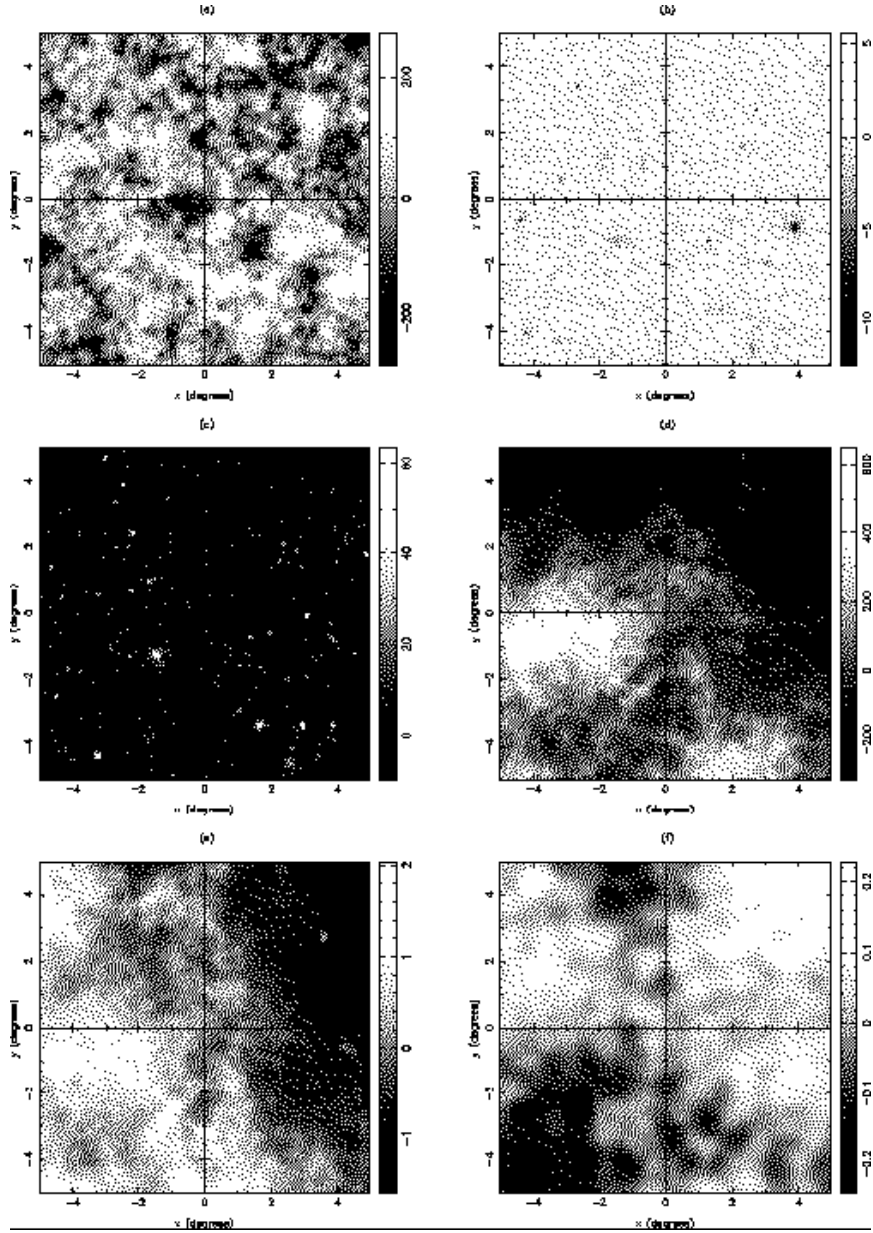, width=4.5in}}
\caption{FIGURE 1. 
The six input components used for the Planck Surveyor
satellite observations. They are shown at 300~GHz in equivalent
thermodynamic temperature. The components shown are a) CMB (CDM realisation), b)
kinetic SZ, c) thermal SZ, d) dust emission, e) free-free emission and
f) synchrotron emission.
}
\end{figure}

We perform two separate analyses to test the effect of relaxing the
assumptions made about the Galactic foregrounds. The first is 
simply to assume the incorrect frequency spectra for the Galactic
foregrounds and test the effect that this has on the
reconstructions. The second is to assume a spatially varying spectral
index for the Galactic foregrounds and try to reconstruct the
components without knowing the form of this spatial
variation. Figure 2 shows the input spatial variation in the
Galactic dust spectral index.

\begin{figure}
\centerline{\psfig{file=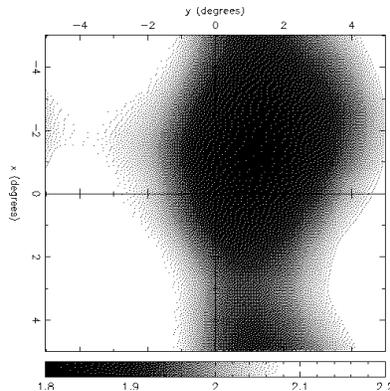, width=2.0in,height=2.0in,angle=270}}
\caption{FIGURE 2.
The spatial variation of the dust spectral index used 
in the simulated Planck Surveyor satellite observations.
}
\end{figure}

{\text 
  \ni 3. RESULTS FROM THE MEM ANALYSIS

We first assume that the Galactic components contributing to the data
do not have a well known spectral dependence. We therefore need to
search over a range of spectral indices to find the best result. This
is done by observing the $\chi^2$ dependence as a function of spectral
index. For the free-free and synchrotron emission it was found that
little difference in the $\chi^2$ value was found when the incorrect
spectral index was used. The reconstructions of the other components
were not altered in this case but the reconstruction of the free-free
and synchrotron were lost. However, in the case of varying both the
dust emissivity and temperature a strong minimum in $\chi^2$ was seen
at the input values (Figure 3). Therefore, it is possible to
fit for the dust emission using the data alone whereas more
information on the free-free and synchrotron are needed to be
confident in their reconstructions.

\begin{figure}
\centerline{\psfig{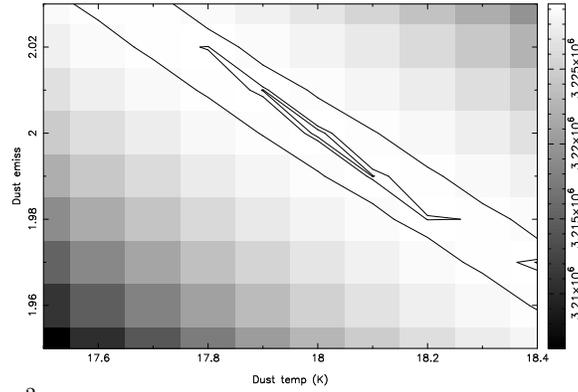}}
\caption{FIGURE 3.
The $\chi^2$ for the reconstructions assuming the temperature
and emissivity of dust in the plot. The input temperature and emissivity were
18K and 2 respectively. The minimum in $\chi^2$ is seen to lie at the
input values.
}
\end{figure}

The errors in each of the reconstructions, if the point at the
$\chi^2$ minimum is used, are not correlated with the
reconstruction itself (see Figure 4) except in the case of a
wrongly assumed spectral index for the free-free and synchrotron where
only the errors on those two components themselves become correlated
with the input signal.

\begin{figure}
\centerline{\psfig{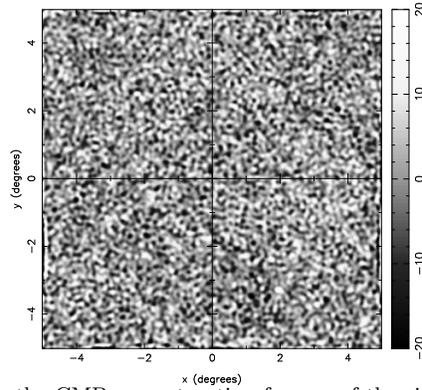}}
\caption{FIGURE 4.
The error in the CMB reconstruction for one of the
simulations. It is seen that the errors are uncorrelated with the
input maps and indeed have no discernible structure themselves.
}
\end{figure}

We now attempt to reconstruct the six channels in the presence of a
spatially varying spectral dependence on the Galactic foregrounds. As
previously the free-free and synchrotron emission are not significant
enough to allow fitting for the spectral index and so a spatially
varying spectrum in these two channels will only worsen their
reconstruction and it is very difficult to fit for this
variation. However, if the dust has a spatially varying spectrum
(as in Figure 2) it is possible to fit for this
variation. The simplest way to do this is to assume that the dust
channel is made up two or more different templates, each with a
separate spectral index which covers the expected range of
variation. For example, we present here the analysis using three dust
templates. The input spatial variation in the dust emissivity was
taken from a Gaussian ensemble with a mean of 2 and a standard
deviation of 0.1. So we use three templates with spectral emissivities
of 2.1, 2.0 and 1.9. Figure 5 shows the reconstruction of the
six channels using this method. As can be seen each of the channels
have been reconstructed very well (the dust templates have been added
together to allow a comparison with the input maps of
Figure 1). Indeed, the error on the CMB reconstruction per
pixel is
still $6\mu$K which is the same as in the case of a single dust
channel with known emissivity (HJLB98). This should be contrasted with
the $10\mu$K error obtained by performing the analysis with only one
dust template (as in HJLB98) assuming a spectral index of 2.0. 
The free-free, synchrotron and thermal SZ have been reconstructed to a
lower amplitude than the input map (this is because most of the
information on these foregrounds occur at the lower frequencies of the
Planck Surveyor which have lower resolutions; see HJLB98). The kinetic
SZ is not reconstructed to a very high significance and only features
associated with strong thermal SZ effects are reconstructed (see
HJLB98).

\begin{figure}
\centerline{\psfig{file=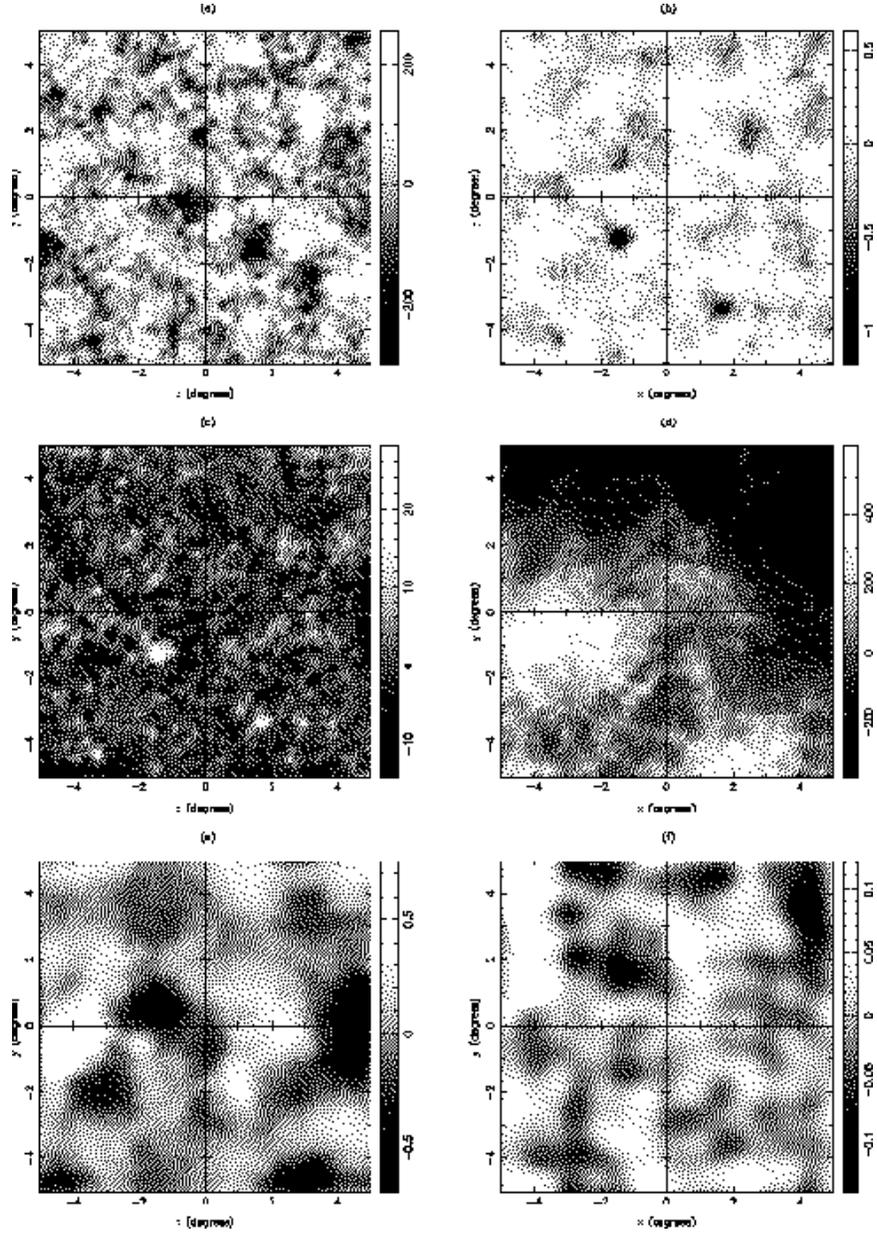, width=4.5in}}
\caption{FIGURE 5.
The reconstruction of the six channels by using three dust
channels with different emissivities to model the spatial variation in
the dust.
}
\end{figure}

  {\text 
  \ni 4. CONCLUSIONS

The maximum entropy method was used to reconstruct data in the
presence of Galactic foregrounds with uncertain spectral
dependencies. It was seen that the Planck Surveyor can fit for a dust
spectral index in the case of a single emissivity and temperature and
can also reconstruct the foregrounds in the presence of a spatially
varying spectral index with little effect on the CMB
reconstruction. It was found that the same was not possible
for the free-free and synchrotron foregrounds but that assuming an
incorrect spectral dependence for these two did not make a significant
difference
to the reconstruction of the other components in the data. 

}

\bsk
\baselineskip = 12pt
{\abstract \ni ACKNOWLEDGMENTS
AWJ acknowledges King's College, Cambridge, for support in the form of
a Research Fellowship.
}

\bsk
\baselineskip = 12pt


{\references \ni REFERENCES
\ssk

\ref 
Bouchet F.R., Gispert R., Boulanger F., Puget J.L., 1997, 
in Bouchet F.R., Gispert R., Guideroni B., Tran Thanh Van J.,eds, 
Proc. 36th Moriond Astrophysics Meeting, 
Editions Fronti\`{e}re, Gif-sur-Yvette, p.~481
\ref
Gispert R., Bouchet F.R., 1997, 
in Bouchet F.R., Gispert R., Guideroni B., Tran Thanh Van J.,eds, 
Proc. 36th Moriond Astrophysics Meeting,
Editions Fronti\`{e}re, Gif-sur-Yvette, p.~503
\ref
Hobson M.P., Jones A.W., Lasenby, A.N., Bouchet, F.R., 1998a, {\sl
MNRAS}, in press
\ref
Hobson M.P., Barreiro R.B., Toffalati L., Lasenby A.N., Sanz J.L., 
Jones A.W., Bouchet F.R., 1998b, {\sl MNRAS}, submitted
\ref
Toffolatti L., Arg\"ueso G\'omez F., De Zotti G., Mazzei P., 
Francheschini A., Danese L., Burigana C., 1998, {\sl MNRAS}, 297, pp 117-127
}                      

\end{document}